# $^{242m}$Am isomer yield in $^{243}$Am(*n*, 2*n*) reaction

V. M. Maslov[1]

220025 Minsk, Byelorussia

*e-mail:mvm2386@yandex.ru*

The reaction $^{243}$Am(*n*, 2*n*) populates either the $T_{1/2}$ =16h ground state $^{242g}$Am with $J^\pi=1^-$ or the $^{242m}$Am isomer state $J^\pi=5^-$ with $T_{1/2}$ =141y. The former state $^{242g}$Am mostly β−decays to $^{242}$Cm, or transmutes to $^{242}$Pu via electron capture. The absolute yield of $^{242g}$Am is compatible with the measured data, estimated by the α–activity of $^{242}$Cm (Norris et al. 1983). The branching ratio defined by the ratio of the populations of the lowest intrinsic states of $^{242}$Am. Calculated yields of ground $^{242g}$Am and isomer $^{242m}$Am states of the residual nuclide $^{242}$Am are used to predict the relative yield of isomer $R(E_n) = \sigma_{n2n}^m(E_n)/(\sigma_{n2n}^g(E_n) + \sigma_{n2n}^m(E_n))$. These populations defined by the γ-decay of the excited states, described by the standard kinetic equation. The ordering of the low and high spin states is different in case of $^{236}$Np and $^{242}$Am nuclides, that explains different shapes of $R(E_n)$ near the (*n*, 2*n*) reaction threshold, though the excitation energy dependences are similar. PFNS data of $^{243}$Am(*n*, *F*) at 4.5 MeV and 14.7 MeV (Drapchinsky et al., 2004, released in 2012) support calculated $^{243}$Am(*n*, *xnf*)$^{1,...x}$ pre-fission neutron contribution to PFNS and calculated exclusive neutron spectra of $^{243}$Am(*n*, 2*n*)$^{1,2}$, feeding the $^{242g}$Am and isomer $^{242m}$Am states.

## INTRODUCTION

The reaction $^{243}$Am(*n*, 2*n*)$^{242m(g)}$Am populates either the $T_{1/2}$ =16h ground state $^{242g}$Am with $J^\pi=1^-$ or the $^{242m}$Am isomer state $J^\pi=5^-$ with $T_{1/2}$ =141y. The former state $^{242g}$Am mostly β−decays to $^{242}$Cm, or transmutes to $^{242}$Pu via electron capture. The yield of the $^{243}$Am(*n*, 2*n*)$^{242g}$Am(β$^-$( ε)$^{242}$Cm($^{242}$Pu) reaction influences the α–activity and neutron activity of the spent fuel due to emerging nuclides $^{242}$Cm and $^{238}$Pu. The yield of the $^{242m}$Am long-lived isomer state, which due to large and odd value of $J^\pi=5^-$ decays to $^{242g}$Am mostly via isomeric transition, gives a path for the $^{244}$Cm build-up via $^{242m}$Am(n, γ)$^{243}$Am(n, γ)$^{244m}$Am(β$^-$(ε))$^{244}$Cm($^{244}$Pu) or $^{242m}$Am(*n*, γ)$^{243}$Am(*n*, γ)$^{244g}$Am(β$^-$) $^{244}$Cm. If not the forbidden β−decay of $^{242m}$Am state, the major path for the $^{244}$Cm build-up would be different. The branching ratio defined by the ratio of the populations of the lowest states of $^{242}$Am. Calculated yields of $^{242g}$Am and isomer $^{242m}$Am states of the residual nuclide $^{242}$Am are used to predict the relative yield of isomer $R(E_n) = \sigma_{n2n}^m(E_n)/(\sigma_{n2n}^g(E_n) + \sigma_{n2n}^m(E_n))$. These populations defined by the γ-decay of the excited states, described by the standard kinetic equation [1, 2]. The absolute yield of $^{242g}$Am is compatible with the measured data, estimated by the α–activity of $^{242}$Cm [3]. The ordering of the low and high spin states is different in case of $^{236}$Np [4, 5] and $^{242}$Am [5], that explains different shapes of $R(E_n)$ near the (*n*, 2*n*) reaction threshold, though their excitation energy dependences are similar.

## BRANCHING RATIO OF SHORT-LIVED $^{242g}$Am(1$^-$) AND LONG-LIVED $^{242M}$Am(5$^-$) STATES IN $^{243}$Am(*n*, 2*n*) REACTION

The approach [1, 2] applied for the modeling branching ratio of the yields of short-lived (1$^-$) and long-lived (6$^-$) of $^{237}$Np(*n*,2*n*) $^{236s(l)}$Np reaction $r(E_n) = \sigma_{n2n}^l(E_n)/\sigma_{n2n}^s(E_n)$ from threshold energy up to 20 MeV allowed to infer the yields of the short-lived state $^{236s}$Np in $^{237}$Np(*n*,2*n*) reaction. The consistent description of the data base on cross sections $^{237}$Np(*n*, *F*), $^{237}$Np(*n*, 2*n*)$^{236s}$Np was achieved [4, 5]. The branching ratio $r(E_n)$ obtained by modeling the nuclide $^{236}$Np levels. Excited levels of $^{236}$Np modeled as predicted Gallher-Moshkowski doublets.

In case of $^{243}$Am(*n*, 2*n*)$^{242m(g)}$Am reaction the branching ratio $r(E_n) = \sigma_{n2n}^m(E_n)/\sigma_{n2n}^g(E_n)$ from threshold energy to 20 MeV could be defined by the ratio of the populations of two lowest states in $^{242}$Am (Fig. 1). These populations are defined by the γ-decay of the excited states, which is described by the kinetic equation [1], and further developed in [2]. The branching ratio $r(E_n)$ is defined by the ratio of the populations of the two lowest states, $^{242g}$Am, with spin $J = 1^-$ and $^{242m}$Am, with spin $J = 5^-$.

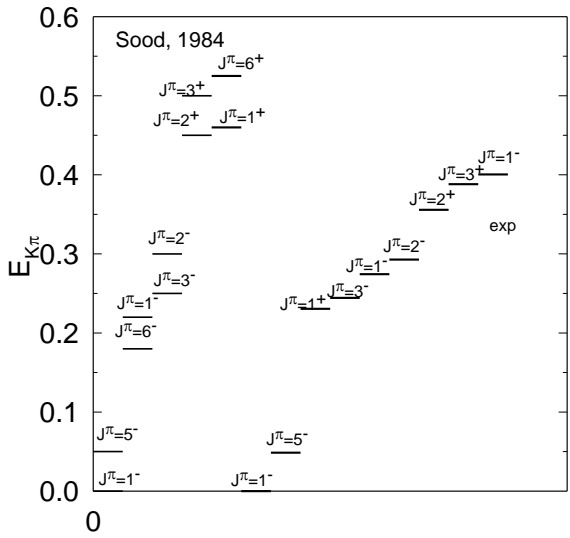
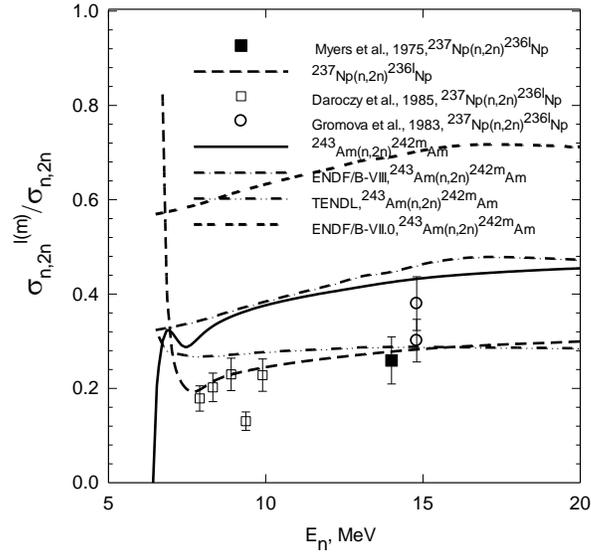

Fig. 1. Levels of $^{242}$Am

Fig.2. Relative yield of long-lived (5$^-$) $^{242m}$Am state in $^{243}$Am($n, 2n$) reaction.

The γ-decay of the excited nucleus described by the kinetic equation [1, 2]:

$$\frac{\partial \omega_k(U,J^\pi,t)}{\partial t} = \sum_{J'\pi'} \int_0^{U_g} \omega_{k-1}(U',J^{\pi'},t) \frac{\Gamma_\gamma(U',J^{\pi'},U,J^\pi)}{\Gamma(U',J^{\pi'})} dt - \omega_k(U,J^\pi,t) \frac{\Gamma_\gamma(U,J^\pi)}{\Gamma(U,J^\pi)}, \quad (1)$$

here $\omega_k(U,J^\pi,t)$ is the population of the state $J^\pi$ at excitation U at time t, after emission of $k$ γ-quanta; $\Gamma_\gamma(U',J^{\pi'},U,J^\pi)$ is the partial width of γ-decay from the $(U',J^{\pi'})$ to the state $(U,J^\pi)$, while $\Gamma(U,J^\pi)$ is the total decay width of the state $(U,J^\pi)$. For any state $(U,J^\pi)$ with the excitation energy 0≤U≤U$_g$, the initial population is

$$\omega_k(U,J^\pi,t=0) = \delta_{ko}\omega_0(U,J^\pi). \quad (2)$$

That equation means that in the initial state we deal with the ensemble of states $(U,J^\pi)$, excited in $^{243}$Am($n, 2n$) reaction. Integrating the Eq. (1) over $t$, one gets the population $W(U,J^\pi)$ of the state $(U,J^\pi)$ after emission of $k$ γ-quanta:

$$\omega_k(U,J^\pi,\infty) - \omega_k(U,J^\pi,0) = \sum_{J'\pi'} \int_U^{U_g} \frac{\Gamma_\gamma(U',J^{\pi'},U,J^\pi)}{\Gamma(U',J^{\pi'})} \int_0^\infty \omega_{k-1}(U',J^{\pi'},t) dt dU' - \frac{\Gamma_\gamma(U,J^\pi)}{\Gamma(U,J^\pi)} \int_0^\infty \omega_k(U,J^\pi,t) dt \quad (3)$$

Denoting the population of the state $(U,J^\pi)$ after emission of $k$ γ-quanta

$$W_k(U,J^\pi) = \frac{\Gamma_\gamma(U,J^\pi)}{\Gamma(U,J^\pi)} \int_0^\infty \omega_k(U,J^\pi,t) dt, \quad (4)$$

and taking into account the condition that $\omega_k(U,J^\pi,\infty)=0$ for any state, belonging to ensemble $(U,J^\pi)$, Eq. (3) could be rewritten as

$$W_k(U,J^\pi) = \sum_{J'\pi'} \int_U^{U_g} \frac{\Gamma_\gamma(U',J^{\pi'},U,J^\pi)}{\Gamma(U',J^{\pi'})} W_{k-1}(U',J^{\pi'})dU' + \omega_k(U,J^\pi,0). \quad (5)$$

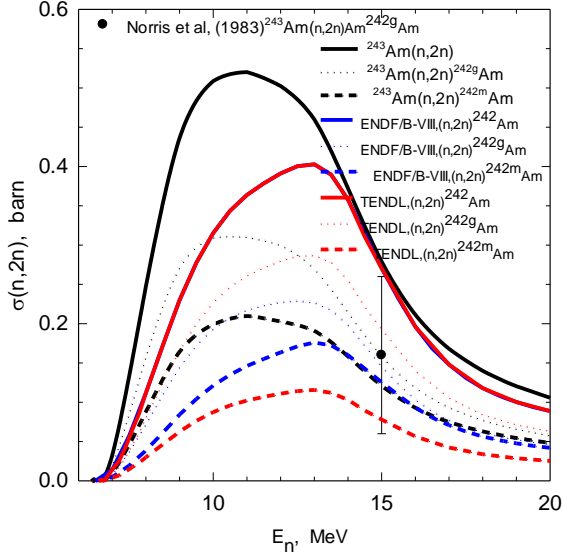 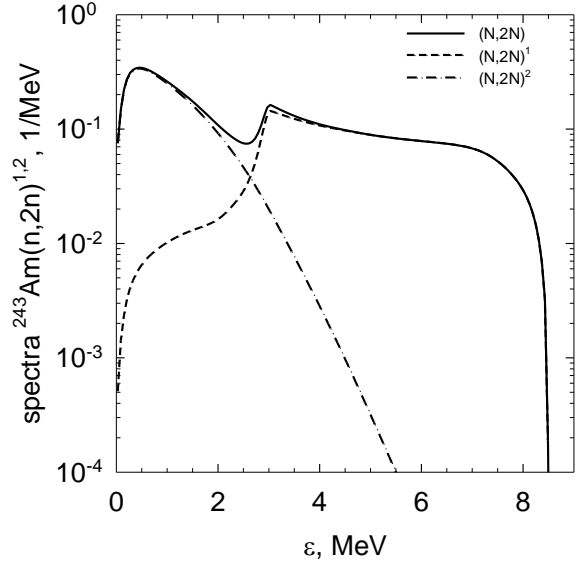

Fig. 3 Cross sections of $^{243}$Am$(n, 2n)$, $^{243}$Am$(n, 2n)$ $^{242m}$Am and $^{243}$Am$(n, 2n)$ $^{242g}$Am.

Fig. 4 Exclusive spectra of $^{243}$Am$(n, 2n)^{1,2}$ at $E_n \sim 15$ MeV

The population of any state $(U, J^\pi)$ after emission of any number of γ-quanta is a lumped sum

$$W(U,J^\pi) = \sum_k W_k(U,J^\pi), \quad (6)$$

then from Eq. (5) one easily gets

$$W(U,J^\pi) = \sum_{J'\pi'} \int_U^{U_g} \frac{\Gamma_\gamma(U',J^{\pi'},U,J^\pi)}{\Gamma(U',J^{\pi'})} W(U',J^{\pi'})dU' + W_0(U,J^\pi). \quad (7)$$

The integral equation (7) in the code STAPRE [9] being solved as a system of linear equations, the integration range $(U,U_g)$ is binned, in the assumption that there are no γ-transitions inside the bins.

The isomer branching ratio depends mostly on the low-lying levels scheme and relevant γ-transitions probabilities. Though experimental data are available for $^{242}$Am [10], we will use a simplified approach, since experimental level scheme and γ-decay intensities are still incomplete. Modeling of low-lying levels of $^{242}$Am in [8] is accomplished based on the assumption that ground and first few excited states are of two-quasi-particle nature. For actinides with quadrupole deformations the superposition principle is usually adopted, the band-head energies of the doubly-odd nucleus are generated by adding to the each unpaired configuration $(\Omega_p, \Omega_n)$, as observed in the isotopic/isotonic (A-1) nucleus, the rotational energy contribution and residual (n−p) interaction energy contribution. The angular momenta of neutron and proton quasi-particles could be parallel or anti-parallel. In the independent quasi-particle model the two-quasi-particle states, $K^+ = |K_n + K_p|$ and $K^- = |K_n - K_p|$, are degenerate. Gallaher-Moshkowski doublets [11] appear because of (n−p) residual interaction. Figure 1 shows employed band-head energies for the two-quasi-particle states expected in the odd-odd nuclide $^{242}$Am up to ~700 keV. The spectroscopic properties of two pairs of proton and neutron single particle states were derived from those experimentally observed in the isotopic (Z=95) and isotonic (N=147) odd-mass nuclei with mass (A-1). Figure 1 shows levels expected, which have similar ordering as experimentally observed [10]. For the band-heads, shown on Fig.1, the rotational bands generated as

$$E_{JK\pi} = E_{JK} + 5.5[J(J+1) - K(K+1)]. \qquad (8)$$

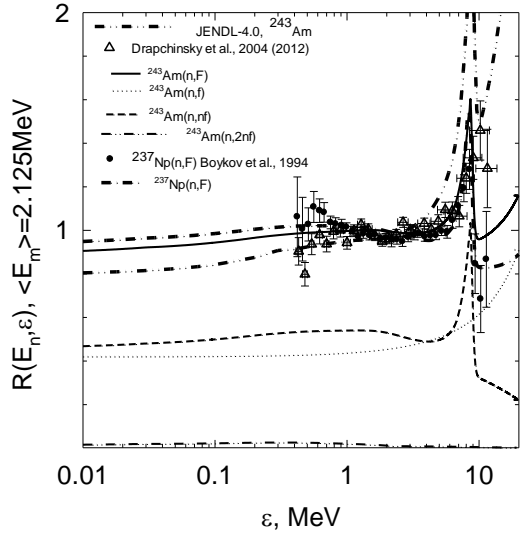
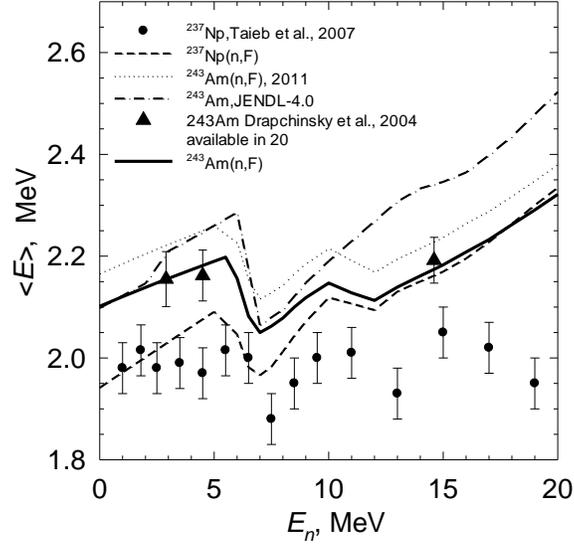

Fig. 5 PFNS of $^{241}$Am(n,F) at $E_n \sim 15$ MeV.   Fig. 6 PFNS $\langle E \rangle$ of $^{243}$Am(n, F)

Obviously, the schema presented on Fig. 1 does not represent a complete set to allow the calculation of absolute yields of $^{242}$Am $(n, 2n)^{242m}$Am and $^{242}$Am $(n, 2n)^{242g}$Am reactions at low $E_n$. Rotational bands were generated up to ~700 keV excitation energy U, modeling levels with spins $J^\pi \leq 10$, in total up to ~70 levels. The simple estimate of the number of levels in odd-odd nuclei as

$$N(U) = e^{2\Delta_0/T}(e^{U/T} - 1), \qquad (9)$$

predicts up to 280 level at $U \sim 700$ keV, $T=0.388$ MeV, $\Delta=12/A^{1/2}$, MeV. We assume that the modeled levels angular momentum distribution would not be much different from realistic estimates. Since the complete data on the γ-transitions are missing, we assumed the simple decay scheme: only E1, E2 and M1 transitions are allowed in a continuum excitation energy range. Inter-band transitions forbidden, i.e., only γ-transitions within the rotational bands are possible. In such approach the populations of the lowest five level doublets could be calculated. Then we assumed that the transition to the isomer state $J^\pi = 5^-$ or low-spin, short-lived ground state $J^\pi = 1^-$ is defined by the "minimal multipolarity" rule. That means the states with spins $J < 3$ should populate the ground state, while those with $J \geq 3$ should feed the isomer state. Then the branching ratio is obtained as the ratio of the populations, derived from Eq. (7):

$$r(E_n) = \frac{\sum_{J \geq (J_l + J_s)/2} W(U, J^\pi)}{\sum_{J < (J_l + J_s)/2} W(U, J^\pi)} \qquad (10)$$

Figure 2 show the relative yield of $^{242m}$Am, calculated for level scheme, presented on Fig. 1. The modeled level scheme appears to be quite compatible with the measured data on the absolute $^{242g}$Am state yield [3] (Fig. 3).

Figure 3 shows, that the yields of the $^{242g}$Am and $^{242m}$Am at $E_n \sim 15$ MeV are still comparable, the latter being lower, as expected for higher spin state in $(n, 2n)$ reaction. In calculation of [5] the ratio is opposite different. The branching ratio for $^{237}$Np$(n, 2n)$ reaction shown on Fig. 2 is much different from that calculated for $^{243}$Am$(n, 2n)^{242m(g)}$Am reaction. It is due to the level spectra differences for residual nuclei $^{236}$Np and $^{242}$Am. Calculated at $E_n \sim 15$ MeV exclusive neutron spectra of $^{243}$Am$(n, 2n)^{1,2}$, feeding the $^{242g}$Am and isomer $^{242m}$Am states, shown on Fig. 4. The main competing neutron channels are $^{243}$Am$(n, nf)^1$ and $^{243}$Am$(n, 2nf)^{1,2}$. The lumped contributions of pre-fission neutrons and respective neutrons coming from fission fragments $^{243}$Am$(n,$

$nf$) and $^{243}$A($n, 2nf$) are on Fig. 5. These partials are consistent with observed prompt fission neutron spectrum of $^{243}$Am($n, F$) [12]. The combined effect of fission chances and exclusive pre-fission neutron spectra leads to the lowering of the average energy $\langle E \rangle$ of the PFNS of $^{243}$Am ($n, F$) near $^{243}$Am ($n, nf$) and $^{243}$Am ($n, 2nf$) reaction thresholds. Similar dips in $\langle E \rangle$ for $^{237}$Np($n, F$) were observed in [14] and interpreted in [4, 5] (Fig. 6).

## CONCLUSIONS

Calculated yields of $^{242g}$Am and isomer $^{242m}$Am states of the residual $^{242}$Am nuclide predict the branching ratio. The branching ratio defined by the ratio of the populations of the lowest states. These populations defined by the γ-decay of the excited states, are described by the standard kinetic equation. The absolute yield of $^{242g}$Am is compatible with the measured data on $^{243}$Am($n, 2n$) $^{242g}$Am, $E_n$~15 MeV [3]. The ordering of the low and high spin states is different in case of $^{236}$Np and $^{242}$Am, that explains different shapes of $r(E_n)$ near the ($n, 2n$) reaction threshold, while excitation energy dependences are similar. PFNS of $^{243}$Am($n, F$) at 14.7 MeV by Drapchinsky (2004, released in 2012) [9] support calculated with the model [15−17] $^{243}$Am($n, xnf$) contribution and exclusive neutron spectra of of $^{243}$Am($n, 2n$)$^{1,2}$, feeding the $^{242g}$Am and isomer $^{242m}$Am states.